\documentclass[review,12pt]{elsarticle}
\usepackage{lineno,hyperref}
\modulolinenumbers[5]
\usepackage{amsmath}
\usepackage{graphicx}
\usepackage{mathtools}

\title{Behaviour of hot electrons under the dc field in chiral carbon 
nanotubes}
\author[els]{M. Amekpewu\corref{cor1}}
\author[rvt]{ K. A. Dompreh}
\author[rvt]{S. Y. Mensah}
\author[ell]{ N. G. Mensah}
\author[rvt]{S. S. Abukari}
\author[els]{R. Musah}
\author[rvt]{A. Twum}
\author[rvt]{R. Edziah}
\address[els]{Department of Applied Physics, University for Development Studies, Navrongo Campus, Navrongo , Ghana}
\address[rvt]{Department of Physics, College of Agriculture and Natural Sciences, U.C.C, Ghana}
\address[ell]{Department of Mathematics, College of Agriculture and Natural Sciences, U.C.C, Ghana}
\cortext[cor1]{Corresponding author\\ Email:} 

\begin{document}
\begin{abstract}
Behaviour of hot electrons under the influence of dc field in 
carbon nanotubes is theoretically considered. The study was done 
semi-classically by solving Boltzmann transport equation with the 
presence of the hot electrons source to derive the current densities. 
Plots of the normalized axial current density versus electric field 
strength of the chiral CNTs reveal a negative differential conductivity 
(NDC). Unlike achiral CNTs, the NDC occurs at a low field about $\mathrm{6\ kV/cm}$ 
for chiral CNT. We further observed that the switch from NDC to PDC 
occurs at lower dc field in chiral CNTs than achiral counterparts. 
Hence the suppression of the unwanted domain instability usually 
associated with NDC and a potential generation of terahertz radiations 
occurs at low electric field for chiral CNTs.  
\end{abstract}

\maketitle
\section*{Introduction}
Carbon nanotubes (CNTs) which are unique tubular structures of nanometric 
diameter with an extremely high length-to-diameter aspect ratio~\cite{1} 
have a wide variety of possible applications~\cite{2, 3, 4}. The primary symmetry 
classification of  carbon nanotubes is as either being achiral (symmorphic) 
or chiral (non-symmorphic)~\cite{5}. Each type of CNT can be either metallic 
or semiconducting depending on their diameter and rolling helicity ~\cite{6}. 
Much progress has been made recently showing that carbon nanotubes (CNTs) 
are advanced quasi-$1$D materials for future high performance 
electronics~\cite{4, 7, 13}. The behaviour of hot electrons in electronic devices has been 
observed since the arrival of the transistor in $1947$~\cite{14}. A number of 
devices have been proposed whose very principle is based on effects of hot 
electrons~\cite{15}. In this paper, we present theoretical framework investigations 
of behaviour of hot electrons under the influence of dc field in chiral carbon 
nanotubes using the semiclassical Boltzmann transport equation. We probe the 
behaviour of the electric current density mainly due to the presence of hot 
electrons in chiral CNTs as a function of the applied dc field along the axis 
of the tube.
\section{Theory}
Suppose  hot electrons is injected axially in a chiral carbon nanotube  which 
is considered as an infinitely long  chains of carbon atoms wrapped along  a 
base helix under the influence of  dc field $E$ . The current densities in 
axial and circumferential directions are calculated by adopting semiclassical 
approximation approach. In the presence of hot electrons source, the motion 
of quasiparticles in dc field is described by Boltzmann transport equation in 
the form~\cite{16, 17}:
\begin{equation}
\frac{\partial f(p)}{\partial t} + v_p \frac{\partial f (p)}{\partial r} eE (t) \frac{\partial f (p)}{\partial p} = - v [f (p) - f_0 (p)] + S(p)
\end{equation}
where $v = 1/\tau,v$  is the scattering frequency of the electrons, 
$\tau$ is relaxation time of electrons, $S(p)$ is the hot electrons source 
function,  $f(p)$  is distribution function, $f_0(p)$ is equilibrium 
distribution function, $\nu_p$ is the electron velocity, $r$ is the electron 
position, $p$  is the electron dynamical momentum, $e$ is the electronic 
charge and t is time elapsed. The energy $\varepsilon(p)$ of the electrons, 
calculated using the tight binding approximation is given as expressed in~\cite{18}  
for a chiral carbon nanotubes:
\begin{equation}
\varepsilon(p) = \varepsilon_0  - \triangle_s \cos(\frac{p_s d_s}{\hbar}) - \triangle_z \cos(\frac{p_z d_z}{\hbar})
\end{equation}
where $\varepsilon_0$ is the energy of an outer-shell electron in an isolated 
carbon atom, $\triangle_z$ and $\triangle_s$ are the real overlapping 
integrals for jumps along the tubular axis and the base helix respectively, 
$p_s$ and $ p_z$ are the components of carrier momentum tangential to the base 
helix and along the tubular axis respectively.
$\hbar = h/2\pi$ and  $h$ is Planck's constant. $d_s$ and $d_z$ are the distances
between the atomic sites $ n1$ and $n1+1$, $n_s$ and $n_s + N_0$ respectively 
along the base helix and the tubular axis where, $(N_0 > 1)$.
The components $v_s$  and $v_z$ of the electron velocity $v_p$ in 
equation ($1$) are respectively calculated from the energy dispersion relation 
equation ($2$) as
\begin{equation}
v_s(p_s) = {\partial \varepsilon (p) \over \partial p_s} = {\triangle_s d_s \over \hbar } \sin ({p_s d_s \over \hbar})
\end{equation}
\begin{equation}
v_z(p_z) = {\partial \varepsilon (p) \over \partial p_s} = {\triangle_z d_z \over \hbar } \sin ({p_z d_z \over \hbar})
\end{equation}
In accordance with~\cite{18,19,20,21}, we find that the distribution function 
is periodic in the quasi-momentum and can be written in Fourier series as:

\begin{eqnarray}
f_0(p_s,p_z) = {n_0 d_s d_z \over 2I_0({\triangle_s \over k_B T})
I_0({\triangle_z \over k_B T}) } \sum^{\infty}_{\alpha = - \infty} 
I_\alpha({\triangle_s \over k_B T})  \sum^{\infty}_{\beta = - \infty} I_\beta
({\triangle_z \over k_B T})\times \nonumber\\
\exp\{ i \alpha p_s d_s / \hbar + \beta p_{zd_z} / \hbar\} 
\end{eqnarray}
\begin{eqnarray}
f_0(p_s, p_z, t) = {n_0 d_s d_z \over 2I_0({\triangle_s \over k_B T}) 
I_0({\triangle_z \over k_B T}) } \sum^{\infty}_{\alpha =- \infty} 
I_\alpha({\triangle_s \over k_B T})  \sum^{\infty}_{\beta =- \infty} 
I_\beta({\triangle_z \over k_B T})\times \nonumber\\
\exp\{ i \alpha p_s d_s / \hbar + \beta p_{zd_z}/\hbar\}\emptyset_\alpha (t)
\end{eqnarray}
where  $f(p_s, p_z, t)$ is the distribution function, and $f_0(p_s, p_z )$ 
is the equilibrium distribution function,  $\emptyset_\alpha(t)$  is the factor 
by which the Fourier transform of the non-equilibrium distribution function 
differs from its equilibrium distribution counterpart, $I_{\alpha,\beta}$  is the 
modified Bessel function of order the ${\alpha,\beta}$, where $\alpha, \beta = 
0, 1, ...$, $ n_0$ is equilibrium particle density and $k_{B}$ is Boltzmann constant.
We consider a hot electron source $S(p)$ of the simplest form given by the 
expression~\cite{17}
\begin{equation}
S(p) = Q \delta(p-p\prime) - {Q \over n_0} f_s (p) 
\end{equation}
where $f_s(p)$ is the stationary solution of Eqn.($1$),  $Q$  is the 
injection rate of hot electrons and $p\prime$ and $p$ are their momentum.
In the case of constant electric field, the solution to Eqn. ($1$) becomes 
\begin{eqnarray}
\emptyset_\alpha(p_s, p_z) = {n_0 d_s d_z \over 2I_0({\triangle_s \over k_B T}) 
I_0({\triangle_z \over k_B T}) } \sum^{\infty}_{\alpha =- \infty} I_\alpha 
({\triangle_s \over k_B T})  \sum^{\infty}_{\beta =- \infty} I_\beta  
({\triangle_z \over k_B T})\times \nonumber\\
{v \over (v + i e \alpha d E)} 
\exp\{i \alpha p_s d_s / \hbar + \beta p_{zd_z} / \hbar\} 
\end{eqnarray}
Therefore Eqn.($6$) becomes
\begin{eqnarray}
f_E(p_s,p_z) = {n_0 d_s d_z \over 2I_0 ({\triangle_s \over k_B T}) I_0 
({\triangle_z \over k_B T})} \sum^{\infty}_{\alpha = - \infty} I_\alpha
({\triangle_s \over k_B T})  \sum^{\infty}_{\beta = - \infty} I_\beta  
({\triangle_z \over k_B T}){v \over (v + i e \alpha d E)}\nonumber\\
 \times \exp\{ i \alpha p_{sd_s} / \hbar + \beta p_{zd_z} / \hbar \}
\end{eqnarray}
The stationary homogeneous distribution function $f_s(p)$ in the presence of 
hot electron source Eqn.($7$) is given by
\begin{equation}
f_s(p) = f_E(p) + f^{\prime}(p)
\end{equation}
Substituting Eqn.($10$) into Eqn.($1$)  we get
\begin{equation}
{\partial f^{\prime} (\varphi) \over \partial \varphi } + ({v \over \Omega} + 
{Q \over n_0 \Omega}) f^{\prime} (\varphi) = {d_s Q \over \Omega} 
\delta(\varphi - \varphi^{\prime}) - {Q \over \Omega n_0} f (\varphi)
\end{equation}
$\Omega =\Omega_{s,z}= {e d_{s, z} E \over \hbar}$ and $d_{s, z}$ is bandwidth,
$\varphi$ and $\varphi\prime$ are the dimensionless momenta of  electrons and 
hot electrons respectively which are expressed as $\varphi=\varphi_{s,z} = d_{s, z}p/\hbar$  
and $\varphi^\prime = \varphi^{\prime}_{s,z}= d_{s, z} p^{\prime}/\hbar$  for  chiral CNTs.
Solving the homogeneous differential equation corresponding to equation ($11$), 
we obtain
\begin{equation}
f^\prime (\varphi ) = C (\varphi) \exp\{-[{v \over \Omega} +{ Q \over n_0 \Omega}] \varphi\}
\end{equation}
Then by differentiating Eqn.($12$), we have
\begin{eqnarray}
{\partial f^\prime(\varphi) \over \partial \varphi} = [{\partial C(\varphi ) \over \partial \varphi} 
- ({v \over \Omega} + {Q \over n_0 \Omega}) C(\varphi)]\times\nonumber\\
\exp\{-[{v \over \Omega} + { Q \over n_0 \Omega}] \varphi\}
\end{eqnarray}
Substituting for $f^\prime (\varphi)$ and ${\partial f^\prime (\varphi)\over 
\partial \varphi}$ in Eqn.($11$), we obtain\\
$${\partial C (\varphi) \over \partial \varphi}  \exp\{-[{v \over  
\Omega} + { Q \over n_0  \Omega}] \varphi\} = {d_s Q \over \Omega} \delta(\varphi - 
\varphi^\prime) - {Q \over \Omega n_0} f_E (\varphi)$$
$$\Rightarrow C (\varphi) = \int \lbrace {d_s p \over \Omega} \delta(\varphi - 
\varphi^\prime) - {Q \over \Omega n_0} f_E (\varphi) \rbrace \exp\{[{v 
\over \Omega} + { Q \over n_0 \Omega}] \varphi\} d \varphi$$
Introducing Dirac-delta transformation
$$\delta (\varphi - \varphi^\prime) = {1 \over 2 \pi} \sum_{\alpha,\beta} \exp\{i (\alpha 
\varphi - \beta \varphi^\prime)\}$$
\begin{equation}
C(\varphi) = {Q  \over 2 \pi \Omega}.{n_0 \over n_0} \int \lbrace \sum_{\alpha,\beta}  
\exp\{i (\alpha \varphi - \beta \varphi^\prime)\} - {Q \over \Omega n_0} f (\varphi)
\rbrace \exp\{[{v \over \Omega} + {Q \over \Omega n_0}] \varphi\}  
d\varphi
\end{equation}
Recall that 
\begin{eqnarray}
f(\varphi) = {n_0 d_s d_z \over 2 I_0 ({\triangle_s \over k_B T}) I_0 
({\triangle_z \over k_B T})} \sum^{\infty}_{\alpha =-\infty}  I_\alpha ({\triangle_s 
\over k_B T})  \sum^{\infty}_{\beta =-\infty}  I_\beta ({\triangle_z \over k_B T})  
{v \over (v + i \exp\{ \beta d_{s,z} E_{s,z})\}}\nonumber\\
\times \exp\{i \alpha d_s d_s /\hbar + i \beta p_z d_z/\hbar\}
\end{eqnarray}
\begin{eqnarray}
C (\varphi) = {Q \over 2 \pi \Omega}.{n_0 \over n_0} \sum_{\alpha}  {\Omega d_s d_z  
\over( i\alpha \Omega + v + {Q\over n_0})} \lbrace \exp\{-i\alpha \varphi\} - {I_{\alpha,\beta} 
({\triangle_s \over kT}, {\triangle_z \over kT}) \over I_0({\triangle_z \over k_B T}) 
I_0({\triangle_s\over k_B T})  ({v\over v + i\alpha \Omega})} \rbrace \nonumber\\ 
\times \exp\{[{v\over \Omega} + {Q \over n_0 \Omega} + i\alpha]\varphi\}
\end{eqnarray}
Substituting Eqn.($14$) into Eqn. ($12$) we get 
\begin{eqnarray}
f^\prime  (\varphi) = {\eta {n_0 d_s d_z} \over 2 \pi \hbar}  \sum_{\alpha} {\Omega 
\exp\{i \alpha \varphi_s\}  \over( i\alpha \Omega + v + \eta \Omega)} \lbrace
\exp\{-i\alpha \varphi^\prime_s\} - {I_{\alpha,\beta} ({\triangle_s \over kT} , {\triangle_z 
\over kT}) \over I_0 ({\triangle_z \over k T}) I_0 ({\triangle_s\over k T})}  
{v \over (v + \alpha l \Omega)} \rbrace 
\end{eqnarray}
Where the nonequilibrium parameter $\eta= {Q\over \Omega n_0}$ and 
$ \eta \Omega = {Q\Omega\over \Omega n_0 }$,
we obtain the general expression for the current density along the tubular 
axis (z-axis) and the base helix (s) as
\begin{eqnarray}
j_{z,s}  = {- \eta n_0 \triangle_{s,z} d^2_{s,z} \over  I_0 ({\triangle_{s,z} 
\over k T})}  \sum_{\alpha}   {\Omega  \over( i\alpha \Omega + v + \eta \Omega)} 
\lbrace \exp\{-i\alpha \varphi^\prime_s\} - {I_{\alpha,\beta} ({\triangle_s \over kT}, 
{\triangle_z \over kT}) \over I_0 ({\triangle_z \over k_BT}) I_0({\triangle_s\over 
k_B T})}  {v \over( v + i\alpha \Omega)} \rbrace 
\end{eqnarray}
\begin{eqnarray}
j_{z}  = {- \eta n_0 \triangle_z d^2_z \over  I_0 ({\triangle_{z} \over k_B T})}  
\sum_{\beta}   {\Omega  \over( i\beta \Omega + v + \eta \Omega)} \lbrace 
\exp\{-i\beta \varphi^\prime_s\} - {I_{\beta} ( {\triangle_z \over kT}) \over 
I_0({\triangle_z \over k_B T})}  {v \over (v + i\alpha \Omega)} \rbrace 
\end{eqnarray}
\begin{eqnarray}
j_{s}  = -{ \eta n_0 \triangle_s d^2_s \over  I_0 ({\triangle_{s} \over k_B T})}  
\sum_{\alpha}   {\Omega  \over( i\alpha \Omega + v + \eta \Omega)} \lbrace 
\exp{-i\alpha \varphi^\prime_s} - {I_{\alpha} ( {\triangle_s \over kT}) \over 
I_0({\triangle_s \over k_B T})}  {v \over (v + i\alpha \Omega)} \rbrace 
\end{eqnarray}
The axial $j_a$ and circumferential $ j_c$ components of the current density 
are given by ~\cite{18, 22}
\begin{eqnarray}
j_a = j_z + j_s \sin \theta_h \\
j_c =  j_s \cos \theta_h 
\end{eqnarray}
where $\theta_h$ is geometric chiral angle. Substituting Eqn.($17$) and 
($18$) into Eqn.($19$) and ($20$) yields
\begin{eqnarray}
j_{a}  = -{ \eta n_0 \triangle_z d^2_z \over  I_0 ({\triangle_{z} \over k_B T})}  
\sum_{\alpha}   {\Omega  \over( i\alpha \Omega + v + \eta \Omega)} \lbrace 
\exp\{-i\alpha \varphi_s\prime\} - {I_{\alpha} ( {\triangle_z \over kT}) \over I_0 
({\triangle_z \over k T})}  {v \over (v + i\alpha \Omega)} \rbrace 
 -{ \eta n_0 \triangle_s d^2_s \over  I_0 ({\triangle_{s} \over k_B T})} \nonumber\\
\sum_{r}   {\Omega  \over( ir \Omega + v + \eta \Omega)} \lbrace
 \exp\{-i\alpha \varphi_s\prime\} - {I_{M} ( {\triangle_s \over kT}) \over I_0 
({\triangle_s \over k_ T})}  {v \over (v + i\alpha \Omega)} \rbrace  
\sin \theta_h
\end{eqnarray}
and
\begin{eqnarray} 
 j_{c} =  -{ \eta n_0 \triangle_s d^2_s \over  I_0 ({\triangle_{s} \over k_B T})}  
\sum_{r}   {\Omega  \over( i\alpha \Omega + v + \eta \Omega)} 
\lbrace \exp\{-i\alpha \varphi_s\prime\} - {I_{\alpha} ( {\triangle_s \over kT}) 
\over I_0 ({\triangle_s \over k T})}  {v \over (v + i\alpha \Omega)} \rbrace  \cos \theta_h
\end{eqnarray}
But from ~\cite{23},
\begin{eqnarray}
 \sin\theta_h = {\sqrt{3} m\over 2\sqrt{n^2+m^2+nm}}
 \end{eqnarray} 
and
\begin{eqnarray}
 \cos\theta_h = {2n+m\over 2\sqrt{n^2+m^2+nm}} 
 \end{eqnarray}
where $n$ and $m$ are integers which denote the number of unit vectors along 
two directions in the hexagonal lattice of graphene. From Eqn. ($21$),
($22$), ($23$) and  ($24$), we finally obtained
\begin{eqnarray}
j_{a}  = -{ \eta n_0 \triangle_z d^2_z \over  I_0 ({\triangle_{z} \over k_B T})}  
\sum_{\alpha}   {\Omega  \over( i\alpha \Omega + v + \eta \Omega)} \lbrace 
\exp\{-i\alpha \varphi^\prime_s\} - {I_{\alpha} ( {\triangle_z \over kT}) \over 
I_0 ({\triangle_z \over k T})}  {v \over (v + i\alpha \Omega)} 
\rbrace \nonumber\\
-{ \eta n_0 \triangle_s d^2_s \over  I_0 ({\triangle_{s} 
\over k_B T})}  \sum_{\alpha}{\Omega  \over( i\alpha \Omega + v + \eta \Omega)} 
\lbrace \exp\{-i\alpha \varphi^\prime_s\} - \nonumber\\
{I_{\beta} ( {\triangle_s \over kT}) \over 
I_0 ({\triangle_s \over k_ T})}{v \over (v + i\alpha \Omega)} \rbrace  
{\sqrt{3} m \over 2 \sqrt{n^2 + m^2 + nm}}
\end{eqnarray}
\begin{eqnarray} 
 j_{c}=  -{ \eta n_0 \triangle_s d^2_s \over  I_0 ({\triangle_{s} \over k_B T})}  
\sum_{\alpha}   {\Omega  \over( i\alpha \Omega + v + \eta \Omega)} \left
\{\exp\{-i\alpha \varphi^\prime_s\} - {I_{\beta} ( {\triangle_s \over kT}) \over I_0 
({\triangle_s \over k T})}  {v \over (v + i\alpha\Omega)}\right\} \nonumber\\
\times{2n + m \over 2 \sqrt{n^2 + m^2 + nm}}
\end{eqnarray}
\section{Results and Discussion}
In figure $1$, we display the behaviour of the normalized axial as well as 
circumferential current density as a function of the electric field for ($4$, $1$) 
chiral CNT injected axially with the hot electrons, represented by the 
none-quilibrium parameter $\eta=1.7 \times10^{-8}$
\begin{figure}[!htp]
\begin{center}
\includegraphics[width=9.5cm]{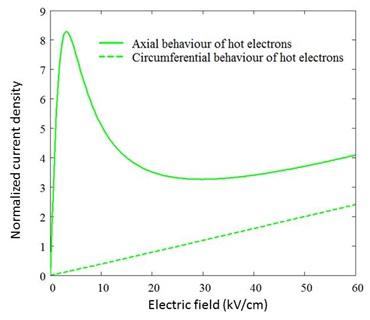}
\caption{A plot of normalized current density versus electric field for 
a ($4,1$) chiral CNT, $\eta=1.7 \times 10^{-8}$,  $v=1 \mathrm{THz}$  
and $T = 287.5\ \mathrm{K}$}
\end{center}
\end{figure}
The normalized axial current density of  ($4$, $1$) chiral CNT exhibits a linear 
monotonic dependence on the applied electric field at lower field (i.e. the 
region of ohmic conductivity). As the electric field increases, the axial 
current density increases and reaches a maximum at about $3.5\ \mathrm{kV/cm}$, and 
drops off, experiencing a negative differential conductivity (NDC)  up to 
about $25\ \mathrm{kV/cm}$ as shown in figure 1. The NDC is due to the increase in the 
collision rate of the energetic electrons with the lattice that induces large 
amplitude oscillation in the lattice, which in-turn increases the scattering 
rate that leads to the decrease in the current~\cite{24}. When the electric field 
exceeds about $25\ \mathrm{kV/cm}$, we once again observe an increase in normalized axial 
current density of chiral carbon nanotube. Hence there is switch from NDC to 
PDC when electric field value of about $25\ \mathrm{kV/cm}$ is exceeded.
This phenomena also occurs when there is an impact ionization. This has been studied 
in superlattices~\cite{28} The physical 
mechanism behind the switch from NDC to PDC has been explained in ~\cite{24, 25}
The normalized circumferential current density of ($4$, $1$) chiral CNT exhibits 
a linear monotonic dependence on the applied electric field as up to $60\ \mathrm{kV/cm}$ 
as shown in figure $1$. Especially from $0$ to about $10\ \mathrm{kV/cm}$, conductivity is 
mainly axially and hence circumferential conductivity is negligible or can 
be ignored. This is due to an extremely high length-to-diameter aspect ratio 
of CNTs up to $132000000:1$ which is significantly larger than any other 
material~\cite{26}. Hence CNTs are quasi-$1$D carbon materials.
In figure $2$, we only display the behaviour of the normalized axial current 
density as a function of the electric field for ($4$, $1$) chiral CNT stimulated 
axially with the hot electrons, represented by the non-equilibrium parameter  
$\eta$. In figure $2$, there is a switch from NDC to PDC  in a ($4$, $1$) chiral CNT 
when $\eta=0.9 \times 10^{-8}$ near $35\ \mathrm{kV/cm}$.
As we increase the rate of hot electrons injection represented by 
non-equilibrium parameter $\eta$  from $0.9 \times 10^{-8}$  to  $3.0 \times 10^{-8}$ 
chiral CNT, the behviour of hot electrons change leading to an increase in 
differential conductivity $\vert{\partial J \over  \partial E}\vert$ as 
well as the peak normalized axial current density as shown in figure $2$
\begin{figure}[!htp]
\begin{center}
\includegraphics[width=9.5cm]{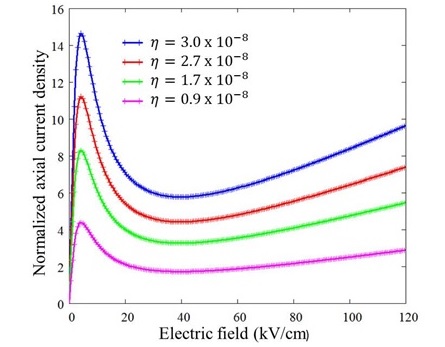}
\caption{Aplot for Normalized axial current density versus Electric field
$\mathrm{(kV/cm)}$ as non-equilibrium parameter $\eta$ increases, $\mathrm{ v = 1THz}$
and $\mathrm{T = 287.5 K}$ } 
\end{center}
\end{figure}
To put the above observations in perspective, we display in figure $3$ a 
3-dimensional behaviour of the normalized axial current density $(Ja)$ 
as a function of the electric field $(E)$ and non-equilibrium parameter 
$(\eta)$. The differential conductivity and the peak of the current density 
are at the lowest values when the non-equilibruim parameter $\eta = 0$. As 
non-equilibrium parameter $\eta$  increases from 0 to $30 \times 10^{-9}$,  both 
differential conductivity and peak current density increase. Therefore, the 
behaviour of hot electrons under the influence of dc field in chiral carbon 
nanotube leads to a switch from NDC to PDC, differential conductivity and the 
peak normalized current density increase.
  \begin{figure}[!htp]
\begin{center}
\includegraphics[width=9.5cm]{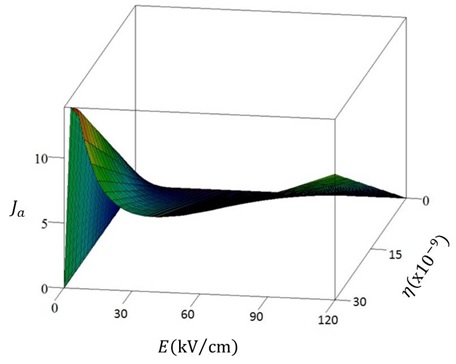}
\caption{ A 3D plot of normalized axial current density $(Ja)$ versus 
electric field $(E)$ and non-equilibrium parameter $(\eta)$} 
\end{center}
\end{figure}          
In figure $4$, we display the behaviour of the normalized axial current 
density as a function of the electric field for the achiral CNTs 
stimulated axially with the hot electrons, represented by the 
non-equilibrium parameter $\eta=1.7 \times 10^{-8}$  as in ~\cite{24} 
\begin{figure}
\begin{center}
\includegraphics[width=9.5cm]{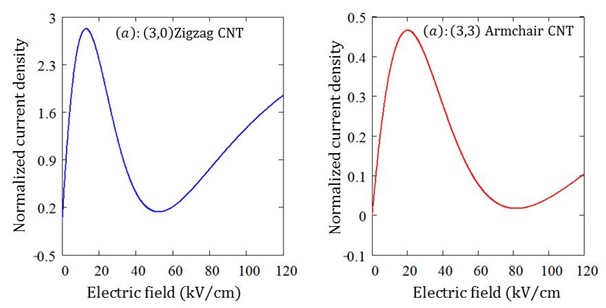}
\caption{A plot of normalized current density versus electric field for (a) 
($3$, $0$) zigzag CNT and (b) ($3$, $3$) 
armchair both stimulated axially with the hot electrons, represented by 
the nonequilibrium parameter  $ \eta=1.7 \times 10^{-8}$ as in~\cite{24}}
\end{center}
\end{figure}
\begin{figure}
\begin{center}
\includegraphics[width=9.5cm]{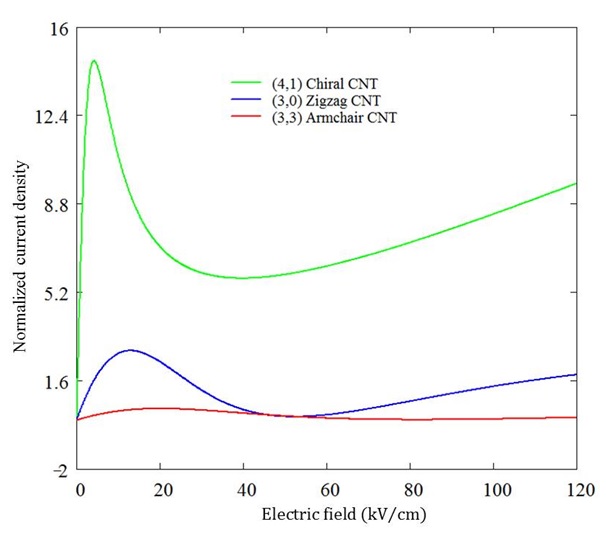}
\caption{A plot of the normalized axial current density versus electric 
field for ($3$, $0$) zigzag CNT, ($3$, $3$) CNT~\cite{24} and ($4$,$1$) CNT stimulated axially 
with the hot electrons, represented by the non-equilibrium parameter 
$\eta = 1.7\times 10^{-8}$}
\end{center}
\end{figure}
We generally observed in figure $5$ that with axial injection of hot electrons 
in ($4$, $1$) chiral CNT, ($3$, $0$) zigzag CNT and ($3$, $3$) armchair CNT under the 
influence of dc field, there is the desirable effect of a switch from NDC to 
PDC characteristics. Thus, the most important tough problem for NDC region 
which is the space charge instabilities can be suppressed due to the switch 
from the NDC behaviour to the PDC behaviour predicting a potential generation 
of terahertz radiations~\cite{25} which have enormous promising applications in 
very different areas of science and technology~\cite{27}. 
However, there are some differences such as differential conductivity, peak 
normalized current density, critical dc field beyond which NDC characteristics 
occurs and finally the value of dc field above which there is a desirable 
switch of NDC to PDC. First and foremost, the differential conductivity 
$\vert{\partial J \over \partial E}\vert$ of ($4$, $1$) chiral CNT is the 
highest, ($3$, $3$) armchair  CNT is the  least while that of ($3$, $0$) zigzag CNT 
is between the two.
Also in figures $4$ and $5$, the peak normalized current density for ($4$, $1$) 
chiral CNT, ($3$, $3$) zigzag CNT and ($3$, $3$) CNT are about $14.20$, $2.70$ and 
$0.46$ respectively. Hence ($4$, $1$) chiral CNT is having the highest peak normalized 
current density follow by ($3$, $0$) zigzag CNT whereas ($3$, $3$) CNT is the least. 
Furthermore, the critical dc fields  that should be exceeded before NDC are 
observed  in ($4$, $1$) chiral CNT, ($3$, $0$) zigzag CNT and ($3$, $3$) armchair CNT are 
about $6\ \mathrm{kV/cm}$, $12\ \mathrm{kV/cm}$ and  $20\ \mathrm{kV/cm}$ 
respectively in figures $4$ and $5$. 
Finally, if the rate of axial injection of hot electrons represented by 
non-equilibrium parameter $\eta=1.7\times 10^{-8}$, there is a switch from NDC 
to PDC near $35\ \mathrm{kV/cm}$,   $50\ \mathrm{kV/cm}$, $75\ \mathrm{kV/cm}$ 
for ($4$, $1$) chiral CNT, ($3$, $0$) 
zigzag  CNT and ($3$, $3$) armchair CNT respectively.

\section{Conclusion}
In summary, behaviour of hot electrons under the influence of dc field in 
chiral carbon nanotubes have been demonstrated theoretically by adopting 
semiclassically approximation approach in solving Boltzmann transport 
equation. It leads to changes in the nature of the dc differential 
conductivity by switching from an NDC characteristics to PDC characteristics 
due to the hot electrons injection. Thus, domain instability can be suppressed, 
suggesting a potential generation of terahertz radiations with enormous 
promising applications in very different areas of  technology, industry and  
research. Unlike achiral CNTs, a potential generation of terahertz radiations 
in chiral counterpart can take place at lower dc field which stems from our 
research findings that the switch from NDC to PDC occurs at relatively low dc 
field.

\end{document}